\documentclass[prl,floatfix,reprint,twocolumn,preprintnumbers,amsmath,amssymb,showpacs,superscriptaddress,longbibliography]{revtex4-2} 

\usepackage[T1]{fontenc}
\usepackage[latin9]{inputenc}
\usepackage{color}
\definecolor{note_fontcolor}{rgb}{0, 0, 0.996094}
\usepackage{amsmath}
\usepackage{amstext}
\usepackage{graphicx}

\def\ll{l}

\makeatletter

\usepackage[dvipsnames]{xcolor}
\usepackage[unicode=true,
 bookmarks=false,
 breaklinks=false,pdfborder={0 0 1},
 colorlinks=false]
 {hyperref}
\hypersetup{
 colorlinks,citecolor=cyan,linkcolor=blue,urlcolor=cyan}

\usepackage{breakurl}

\makeatother

\usepackage{babel}
\begin{document}
\title{Signatures of Chaos in Non-integrable Models of Quantum Field Theory}
\author{Miha Srdin\v{s}ek}
\email[]{miha.srdinsek@upmc.fr}
\author{Toma\v{z} Prosen}
\author{Spyros Sotiriadis} 
\affiliation{Faculty of Mathematics and Physics, University of Ljubljana, SI 1000 Ljubljana, Slovenia}

\begin{abstract}
We study signatures of quantum chaos in (1+1)D Quantum Field Theory (QFT) models. 
Our analysis is based on the method of Hamiltonian truncation, a numerical approach for the construction of low-energy spectra and eigenstates of QFTs that can be considered as perturbations of exactly solvable models. 
We focus on the double sine-Gordon, also studying the massive sine-Gordon and ${\phi^4}$ model, all of which are non-integrable and can be studied by this method 
with sufficiently high precision from small to intermediate perturbation strength. 
We analyse the statistics of level spacings and of eigenvector components,  
which are expected to follow Random Matrix Theory  predictions.
While level spacing statistics are close to the Gaussian Orthogonal Ensemble as expected, 
on the contrary, the eigenvector components follow a distribution markedly different from the expected Gaussian. 
Unlike in the typical quantum chaos scenario, the transition of level spacing statistics 
to chaotic behaviour takes place already in the perturbative regime. 
Moreover, the distribution of eigenvector components does not appear to change or approach 
Gaussian behaviour, even for relatively large perturbations. 
Our results suggest that these features are independent of the choice of model and basis. 
\end{abstract}
\maketitle

\emph{Introduction.}{---} The physics of non-integrable quantum systems has been 
successfully described by quantum chaos theory, 
which states that their spectral statistics are given by Random Matrix Theory (RMT) 
i.e. exhibit the same behaviour as 
matrices whose elements are randomly chosen from a Gaussian distribution.
These conjectures \cite{casati,bohigas,berry1977} have been verified for a broad class of single-particle models,
where they have been explained in terms of semi-classical periodic orbit theory \cite{bogomolny1988,haake2004}. 
More recently research focus has shifted to many-body systems \cite{Santos-Rigol,Santos-Rigol2,Santos-Rigol3,D_Alessio_2016,Alba,Haque2,Haque1,Haque3,Haque2020,Foini2019,DeLuca}, 
where RMT predictions have been verified numerically and in certain cases even analytically 
\cite{kos2018,bertini2018}. 
Chaoticity tests in quantum many-body models are, however, almost exclusively limited to discrete (lattice) models, 
leaving continuous models unexplored.
Among them, relativistic Quantum Field Theories (QFTs) and their dynamics 
lie at the cornerstone of important open questions of theoretical physics, 
like the black hole information paradox \cite{Hayden_2007}, 
making the study of ergodicity and chaos in QFT a topic of fundamental interest.

{Significant  
progress in this direction has been made 
based on new theoretical concepts and indicators 
\cite{Sekino2008, Shenker2014, Maldacena:2016aa}. 
Nevertheless, the emergence of quantum chaos in QFT remains
poorly understood in terms of  
the more traditional measures of level spacing and eigenvector  
statistics \cite{Haake_2010}. 
Studying level spacing statistics is the best way of detecting level repulsion, 
the characteristic property of random matrix spectra. 
On the other hand, a Gaussian distribution of eigenvector components is 
an important indication of validity of the Eigenstate Thermalisation Hypothesis (ETH) \cite{Deutsch91,Srednicki99}, which explains how thermalisation emerges from the 
dynamics of non-integrable quantum systems \cite{Hortikar1998}.}

\begin{figure}[t]
\includegraphics[width=1\columnwidth]{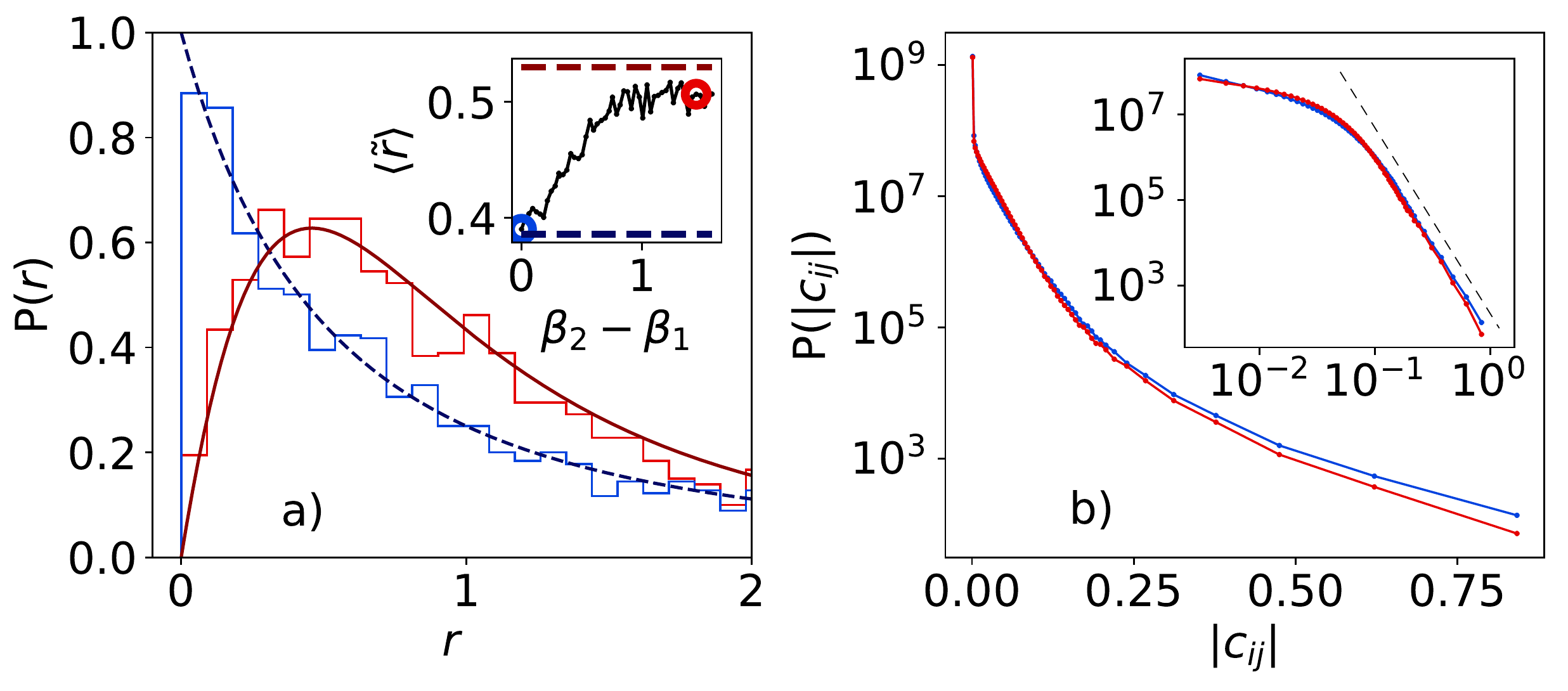}
\caption{Statistics of level spacings and eigenvector components in the double sine-Gordon (non-integrable)
and sine-Gordon (integrable) model. 
(a) Change in the {${r}$ distribution} of DSG from
${(\beta_{1},\beta_{2})=(2.5,2.5)}$ (integrable SG point) to ${(\beta_{1},\beta_{2})=(1.0,2.5)}$ (non-integrable point)
 compared to the predictions for integrable models (dashed blue curve) and to the RMT predictions (red curve), respectively. 
 Inset shows change in the average of ${\tilde{r}}$ when varying ${\beta_1}$. 
(b)  Distribution of the absolute values of eigenvector components ${|c_{ij}|}$ {for the same two points}, in log scale and log-log scale (inset). The GOE distribution of ${|c_{ij}|}$ is Gaussian. 
Instead, we observe that, although the level spacing statistics of DSG is 
GOE-like, the statistics of ${|c_{ij}|}$ is not.}
\label{fig:1}
\end{figure}

The main obstacle in performing chaoticity tests in QFT is that, unlike
for lattice models of condensed matter physics, QFT models are continuous
and thus live in an infinite dimensional Hilbert space. 
Therefore, exact computation of 
energy spectra is not an option 
for non-integrable models,
and we inevitably resort to approximate numerical methods. 
The challenge is then to achieve sufficiently high accuracy in a sufficiently
large part of the spectrum, so that a statistical analysis is possible and reliable. 
An ideal method for this task is the ``Truncated
Conformal Space Approach'' (TCSA) \cite{YUROV1990,YUROV1991,TCSA-sG1}, 
{more generally the Hamiltonian truncation method \cite{Hogervorst2014,James2017}.} 
The TCSA is 
based on the algebraic toolkit of
Conformal Field Theory (CFT) and insights from Renormalisation Group
theory, which can capture efficiently non-perturbative effects {in the low-energy spectrum}, 
and is especially suitable for (1+1)D models. 
A pioneering study of quantum chaos indicators using this method was presented
in \cite{Brandino_2010} for the tricritical and tetracritical Ising field theories, 
demonstrating that their level spacing statistics agree with 
the theoretical expectations in both the integrable and non-integrable case, 
and observing their crossover for varying parameters.

In this Letter we study two 
independent and equally important signatures of quantum chaos, the
distribution of level spacings as quantified by the consecutive level spacing 
ratios ${\tilde{r}}$ \cite{Oganesyan_2007,Atas2013} and the distribution
of eigenvector components. We study a class of (1+1)D
models: the sine-Gordon model (SG), which is integrable, and the double
sine-Gordon (DSG), massive sine-Gordon (MSG) a.k.a. 
 Schwinger--Thirring, and ${\phi^4}$ model, which are all non-integrable. 
We verify that level spacings follow the {expected} 
Poisson distribution for SG
and GOE distribution for DSG, MSG and ${\phi^4}$ model 
to a very good approximation. 
{GOE behaviour is {actually} observed already in the weakly perturbed CFT regime, 
in contrast to what typically happens in single-particle models.} 
Surprisingly, we find that, even when the level spacing distribution is close
to GOE, the eigenvector component distribution is markedly different
from the Gaussian  found in RMT \cite{Brody1981} (Fig.~\ref{fig:1}). 
On the contrary, it exhibits at best exponential scaling followed by an algebraically decaying tail, which contradicts the RMT prediction. This last feature is 
robust and independent of the model and parameter values. 
We validate our observations by pushing the limits of TCSA's potential to achieve high accuracy 
and devising a reliable measure of truncation error, which is 
crucial for distinguishing physical behaviour from numerical artefacts.

\emph{Models, Method \& Observables.}{---} We consider 
the following models: the SG with Hamiltonian ${H_{SG}=H_{0}+\lambda V_{\beta}}$, 
the DSG  ${H_{DSG}=H_{0}+\lambda_{1}V_{\beta_{1}}+\lambda_{2}V_{\beta_{2}}}$, the MSG  ${H_{MSG}=H_{0}+\lambda V_{\beta_1}+m^{2}U_{2}}$, and the ${\phi^4}$ model  ${H_{\phi^4}=H_{0}+m^{2}U_{2} + \lambda U_{4}}$, 
where 
\begin{eqnarray}
& H_{0} =\tfrac{1}{2}\textstyle\int [\pi^2 -(\partial_x\phi)^{2}]\,\mathrm{d}x,  \\ 
&U_{n} =\tfrac{1}{n!}\textstyle\int\phi^{n}\,\mathrm{d}x,\quad
V_{\beta} =-\textstyle\int\cos\beta\phi\,\mathrm{d}x.\nonumber
\end{eqnarray}
The SG is a prototypical integrable QFT possessing
topological excitations 
\cite{ZAMOLODCHIKOV1995,TCSA-sG1,TCSA-sG2,TCSA-sG3}
and is equivalent to the massive Thirring model 
\cite{Coleman,Mandelstam}. It has applications in condensed
matter and atomic physics \cite{Giamarchi2004} and has been simulated experimentally 
\cite{exp-sG,Schweigler2}. The DSG is non-integrable and 
also topologically non-trivial 
\cite{Delfino_1998,Bajnok_2001,Mussardo_2004,Tak_cs_2006}. Lastly,
the MSG 
is equivalent to the 
Schwinger--Thirring model, 
reducing to (1+1)D QED 
at ${\beta=\sqrt{4\pi}}$ \cite{Coleman_1975,Coleman_1976}.

All the above models can be seen as perturbations of the 
free boson CFT ${H_{0}}$ by 
relevant operators ${V}$ and as such they
can be studied using TCSA. This method yields numerical approximations
of the low-energy spectrum of ${H=H_{0}+\lambda V}$ 
based on the 
simple idea of computing the matrix elements of ${V}$ 
in an energy-truncated basis ${\{|\Phi_{n}^{0}\rangle: \, E_n^0\leq E_\text{cut}\}}$ 
of ${H_{0}}$ and diagonalising the resulting finite matrix approximation of ${H}$. 
If ${V}$ does not couple significantly the low- with the high-energy spectrum of ${H_{0}}$, 
which is true for relevant perturbations, 
then the numerical spectrum is expected to converge to the exact upon increasing the truncation cutoff ${E_{\text{cut}}}$. 
TCSA has been successfully applied to the SG
\cite{TCSA-sG1,TCSA-sG2,TCSA-sG3,KST,Kukuljan_2020}, 
DSG \cite{Bajnok_2001,Tak_cs_2006} and recently also the Schwinger model, a special limit of MSG \cite{Kukuljan_2021}, 
while a similar Hamiltonian truncation method 
has been used for the ${\phi^4}$ model  \cite{Hogervorst2014,Rychkov2014,Rychkov2015,Bajnok2016,Elias-Miro2017}. 

Using TCSA we compute a low-energy part of the spectra ${E_{n}}$
and eigenvectors ${|\Phi_{n}\rangle}$ of the above models for 
various parameter values and analyse their statistics. More specifically,
we compute the distributions of level spacings ${s_{n}=E_{n+1}-E_{n}}$, 
of consecutive level ratios ${r_{n}}$ and ${\tilde{r}_{n}}$ defined
as \cite{Oganesyan_2007}
\begin{equation}
{r}_{n}={s_{n}}/{s_{n-1}}, \quad \tilde{r}_{n}=\min\left({r}_{n},1/{r}_{n}\right)
\end{equation}
and of eigenvector components 
${c_{ij}=\langle\Phi_{i}^{0}|\Phi_{j}\rangle}$ in the TCSA basis. 
Since all models are time-reversal symmetric, the corresponding RMT ensemble is the GOE where {the ${r}$-distribution is ${P_{GOE}(r)\propto ({r+r^2})/{(1+r+r^2)^{5/2}}}$  
and that of ${\tilde{r}}$ is the restriction of the latter to the interval ${[0,1]}$, with 
mean value ${\langle\tilde{r}\rangle_{GOE}\approx 0.536}$ \cite{Atas2013}}. 
Conversely, in integrable models level spacings follow the Poisson distribution \cite{BerryTabor} with ${\langle\tilde{r}\rangle_{P}\approx 0.386}$. Compared to other tests of level spacing statistics, ${\tilde{r}}$ has the advantage 
of being independent of the local 
level density, therefore no `unfolding' \cite{Haake_2010} 
is necessary. 
For the eigenvector component distribution, the RMT prediction is Gaussian, resulting in the 
Porter--Thomas distribution for their absolute values \cite{Brody1981}, while for integrable models it is expected to be algebraic \cite{Samajdar2018}.

{To minimise numerical errors we use truncated bases much larger than in previous studies (${\sim85000}$ states at the highest cutoff). Moreover, to ensure our results are sufficiently accurate, we verify convergence using rather strict truncation error estimates, based on measures of their correlations at successive cutoffs (see figure captions and Supp. Mat. \cite{SM}).}

\begin{figure}[ht]
\includegraphics[width=1\columnwidth]{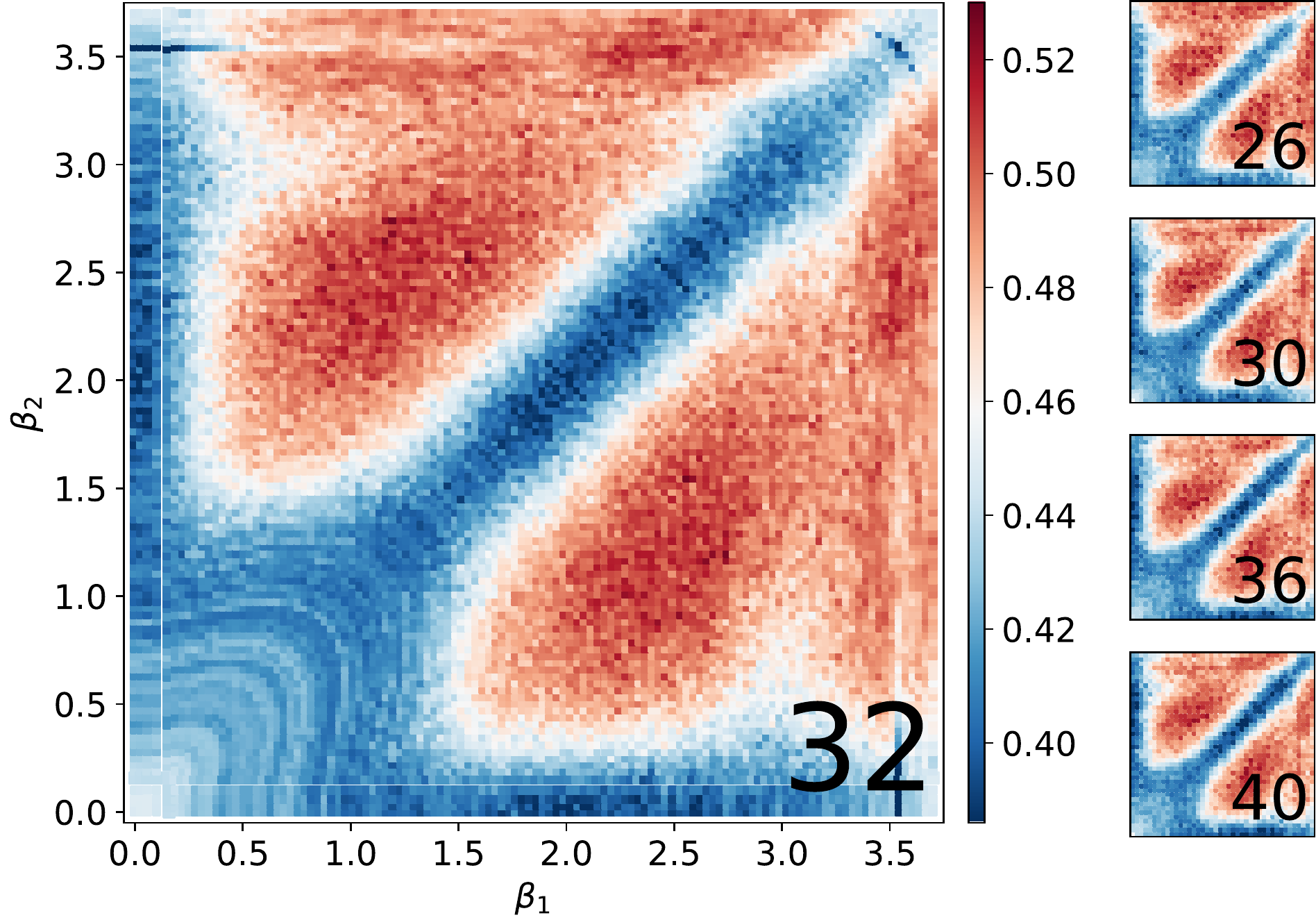}
\caption{
Density plot of ${\langle\tilde{r}\rangle}$ in DSG as a function of 
${\beta_{1}}$
and ${\beta_{2}}$. The model is non-integrable with the exception of
the three lines ${\beta_{1}=0,\beta_{2}=0}$ and ${\beta_{1}=\beta_{2}}$,
where it reduces to the SG. The 
${\langle\tilde{r}\rangle}$ is indeed close to ${\langle\tilde{r}\rangle_{P}}$ (dark blue) along the SG lines and in their vicinity, while it approaches 
${\langle\tilde{r}\rangle_{GOE}}$ (dark red) away from them. 
Plots for different truncation cutoffs ${E_\text{cut}}$ are included for comparison (\emph{right}). 
(Parameters: ${\ll_1=\ll_2=1}$, energy window: 1000--3000 levels, ${E_\text{cut}}$ reported 
at the bottom right corner of each plot).
}
\label{fig:2}
\end{figure}

\begin{figure}[t]
\includegraphics[width=\columnwidth]{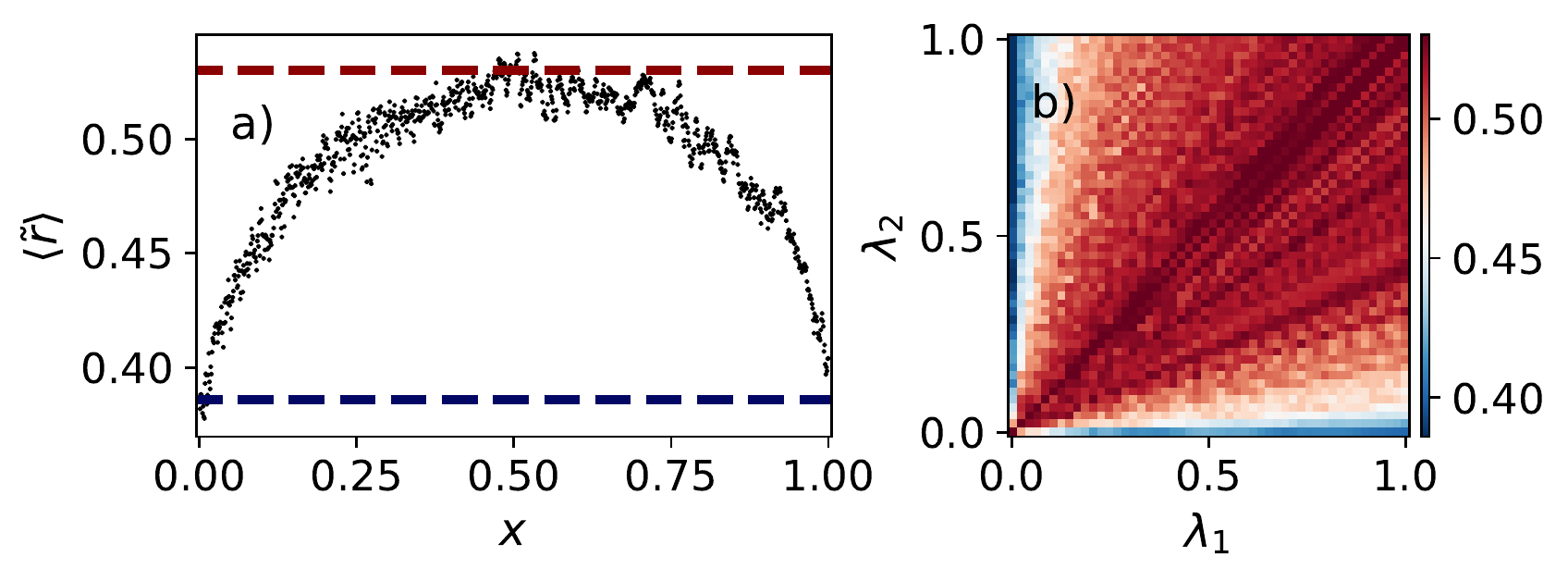}
\caption{
Dependence of ${\langle\tilde{r}\rangle}$ on the perturbation strength in DSG.  
(a) Plot of ${\langle\tilde{r}\rangle}$ for ${\lambda_{1},\lambda_{2}\to0}$ as a function of the mixing ratio ${x=\lambda_{1}/(\lambda_{1}+\lambda_{2})}$. 
(b) Density plot of ${\langle\tilde{r}\rangle}$ as a function of ${\lambda_{1}}$ and ${\lambda_{2}}$. 
The 
SG lines 
correspond to ${\lambda_{1}=0}$ and ${\lambda_{2}=0}$. 
Note that ${\langle\tilde{r}\rangle \approx \langle\tilde{r}\rangle_{GOE}}$  
even in the immediate vicinity 
of the unperturbed CFT model. 
(Parameters: ${(\beta_{1},\beta_{2})=(1,2.5)}$). 
}
\label{fig:2b}
\end{figure}

\begin{figure*}[ht]
\includegraphics[width=1\textwidth]{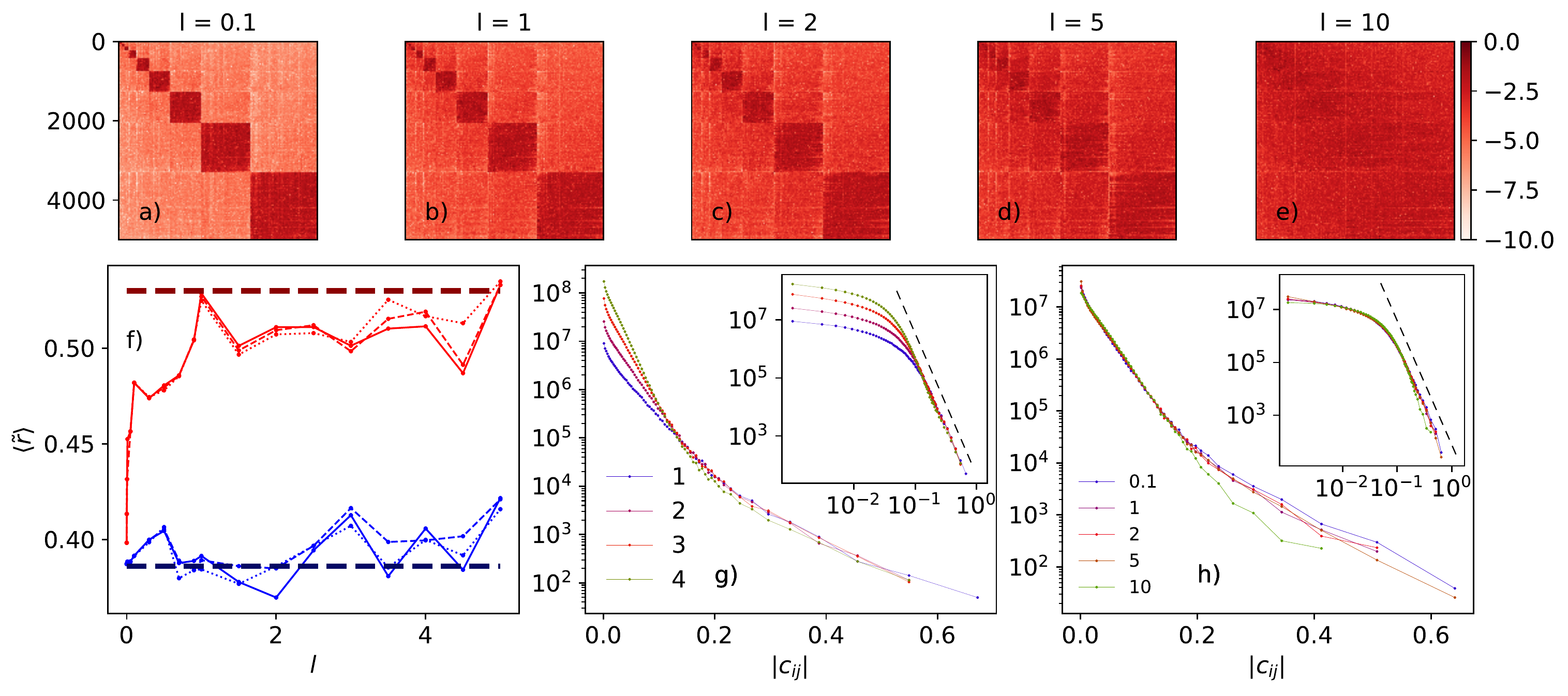}
\caption{
Dependence of spectral properties on perturbation strength. 
(a--e) Matrix plot of the eigenvector matrix ${|c_{ij}|}$ of DSG at 
 different ${\ll=\ll_1=\ll_2}$, in logarithmic scale. 
  {Only the low-energy part (top left corner) of the 
  	matrix is shown, which is fully convergent for ${\ll\leq 2}$, partially convergent for ${\ll\approx5}$ (lowest ${\sim 3000}$ levels convergent), and poorly convergent for ${\ll\approx10}$. }
 (f) Plot of ${\langle\tilde{r}\rangle}$
as a function of ${l}$ at different cutoffs (dotted: 38, dashed: 40, solid line: 42) for DSG (red) and SG (blue). 
(g) Distributions of ${|c_{ij}|}$ 
in the last four boxes shown in (a--e), labelled as 1--4. 
(h)  Distribution of ${|c_{ij}|}$ in the third 
largest box for different ${l}$. 
(Parameters: ${(\beta_{1},\beta_{2})=(1.0,2.5)}$ for DSG, ${\beta=2.5}$ for  SG.) }
\label{fig:3}
\end{figure*}

\emph{Level spacing statistics.}{---} We start by analysing the statistics of ${r}$ values. 
Fig.~\ref{fig:1}.a shows the distribution ${P(r)}$ for DSG 
at two different choices of parameter values, one integrable ${(\beta_{1},\beta_{2})=(2.5,2.5)}$
(SG) and one non-integrable ${(\beta_{1},\beta_{2})=(1.0,2.5)}$. 
The parameters ${\lambda_{1},\lambda_{2}}$ have been chosen so that the energy gap between the ground and first excited state is of 
the same order as the inverse system size ${L^{-1}}$ 
(more precisely, ${\ll_1=\ll_2=1}$ where ${\ll_i=m_{\beta_{i}}L}$ and ${m_{\beta}}$ is the SG breather mass \cite{SM}). These values are within the perturbative regime where convergence is optimal. 
We observe that the two distributions agree quite well with the Poisson and GOE distributions,
respectively. The change of statistics 
can be demonstrated by the mean value ${\langle\tilde{r}\rangle}$
for varying ${\beta_{1}}$ at fixed ${\beta_{2}}$ (Fig.~\ref{fig:1}.a, 
inset). Starting from the Poisson value for ${\beta_{1}=0}$, ${\langle\tilde{r}\rangle}$
increases towards the GOE value, fluctuating close and below it.
The complete dependence of ${\langle\tilde{r}\rangle}$ on both 
${\beta_{1}}$ and ${\beta_{2}}$ is shown in Fig.~\ref{fig:2} in the form of a ``phase
diagram''. 
The ${\langle\tilde{r}\rangle}$ is close to Poisson 
along the SG lines (${\beta_1=0, \beta_2=0}$ and ${\beta_1=\beta_2}$), whereas 
it approaches ${\langle\tilde{r}\rangle_{GOE}}$ in the areas away from these lines.

By independently varying the perturbation strength parameters ${\lambda_1,\lambda_2}$ 
at fixed ${(\beta_{1},\beta_{2})=(1,2.5)}$ 
({Fig.~\ref{fig:2b}}) we find that ${\langle\tilde{r}\rangle}$ is close to ${\langle\tilde{r}\rangle_{GOE}}$ 
even in the immediate vicinity of the unperturbed CFT model, 
i.e. for ${\lambda_1,\lambda_2\to 0}$, 
as long as the ratio ${\lambda_1/\lambda_2}$ is kept fixed at ${\sim 1}$. 
This is somewhat surprising given that chaoticity 
typically emerges far from integrable points and outside the perturbative regime. 
Chaotic ${\langle\tilde{r}\rangle}$ values can indeed be verified from first-order perturbation theory results 
({Fig.~\ref{fig:2b}}.a) \cite{SM}. 
 The fluctuations of ${\langle\tilde{r}\rangle}$ can be partially attributed 
 to the relatively small energy window (2000 levels), 
 as random matrices of the same size display similar fluctuations. 
 Nevertheless, 
we notice that ${\langle\tilde{r}\rangle}$ is predominantly below 
 ${\langle\tilde{r}\rangle_{GOE}}$, meaning that this is still 
 a transitional, not completely chaotic behaviour.

The differences between ${\langle\tilde{r}\rangle}$ at different 
cutoffs (Fig.~\ref{fig:2}) are negligible,  
with the best convergence achieved for small ${\beta_1,\beta_2}$. 
However, even if ${\langle\tilde{r}\rangle}$ converges at some
cutoff to ${\langle\tilde{r}\rangle_{GOE}}$, this does not necessarily mean that this
is the correct physical value, since a non-convergent spectrum is
also likely to be RMT-like. 
For this reason, we check the convergence using an error estimate based
on the averaged absolute differences of ${\tilde{r}}$ values between successive
cutoffs and verifying that the error decreases with increasing cutoff \cite{SM}. 
We empirically find that increasing ${\beta_{i}}$ or ${\ll_i}$ 
results in larger truncation errors, making the numerical data less reliable. 
Moreover, in TCSA convergence is achieved in the lowest part of the computed spectra, 
with the truncation effects 
increasing at higher levels. 
For the parameters of Fig.~\ref{fig:1}, a sufficiently good level of convergence of ${r}$ values 
is achieved for the lowest ${\sim3000}$ levels at ${E_\text{cut}=42}$ 
(in units ${\varepsilon=\pi/L}$).

\emph{Eigenvector statistics.}{---} Let us focus on the statistics 
of eigenvector components ${c_{ij}}$ in DSG. 
Fig.~\ref{fig:1}.b shows 
the distribution of their absolute values in log scale for the same choice of parameters 
as in Fig.~\ref{fig:1}.a, one exhibiting Poisson and the other GOE level spacing statistics. 
Despite the clear difference in the latter, 
the eigenvector distributions 
are practically the same in both cases and 
different from the Gaussian prediction of RMT. 
In the bulk of the distribution the scaling is at best exponential 
while the tails decay slower, 
like an algebraic function. This is in strong contrast with theoretical expectations for chaotic models \cite{Brody1981}. 
To eliminate truncation effects, we have again restricted the analysis to the convergent low-energy part of the matrix ${c_{ij}}$.

To gain a deeper insight into this observation, we look more closely into the structure of the 
matrix ${c_{ij}}$. 
Fig.~\ref{fig:3}{.a--e} shows ${c_{ij}}$ for DSG at increasing perturbation strength 
 ${\ll=\ll_1=\ll_2}$. 
We observe that for small ${\ll}$, 
${c_{ij}}$ is characterised by 
an approximately block-diagonal form, 
which is easily explained by perturbation theory 
given that the CFT spectrum is organised in degenerate energy shells \cite{SM}. 
For increasing ${\ll}$ this block structure fades away and ${c_{ij}}$ becomes more uniform, 
even though a pattern of fine structure remains always visible.

\begin{figure}[htbp]
\includegraphics[width=1\columnwidth]{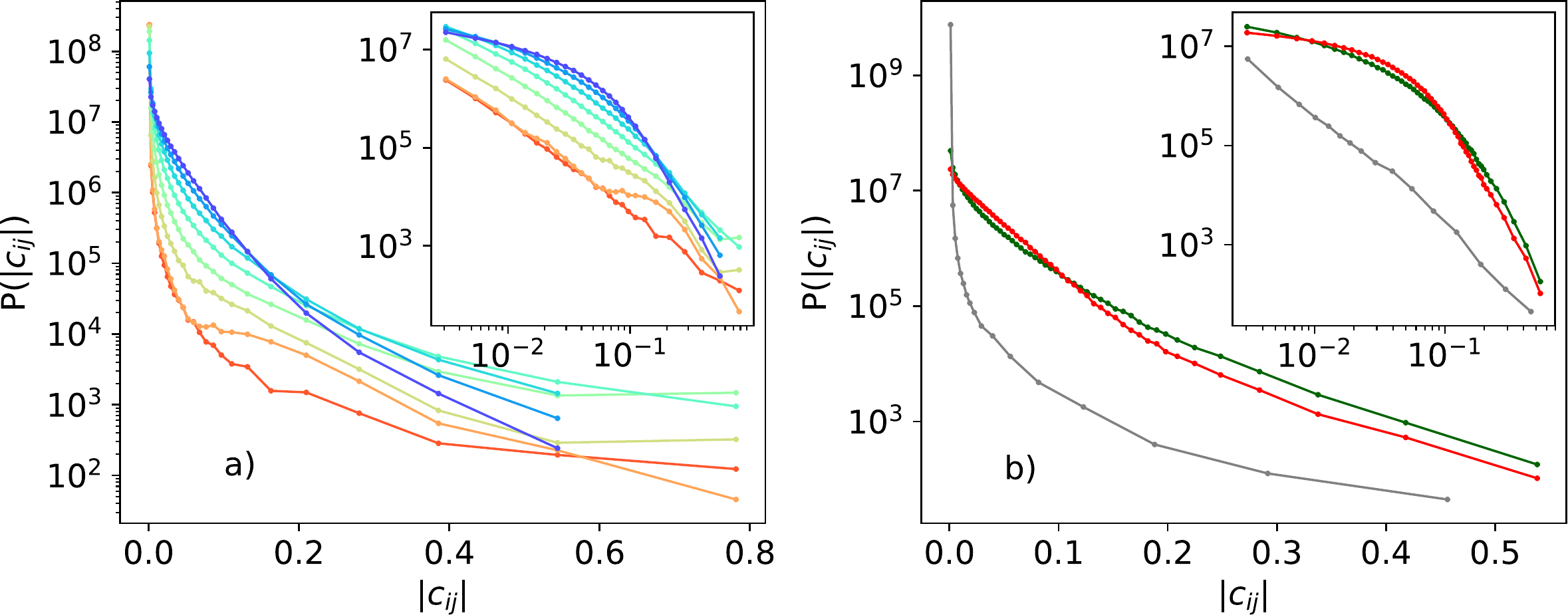}
\caption{
(a) Distribution of ${|c_{ij}|}$ in SG in log and log-log
 scale (inset) 
at different 
 ${\beta}$ from ${0}$ (red)  to ${2.5}$ (blue). The 
 exponential scaling in the bulk is not a special property of the non-integrable 
 DSG but is also present in SG. 
 (b) Comparison of ${|c_{ij}|}$ distributions in DSG (red), MSG (green) and ${\phi^4}$ model (grey), 
 always in windows with GOE spectral statistics. Even
 though the ${\phi^4}$ model is expanded in a different and 
 less exceptional basis (KG), the distribution is still not GOE-like but clearly algebraic. 
(Model parameters, ${\langle\tilde{r}\rangle}$ values: 
 DSG: ${(\beta_{1},\beta_{2})=(1.0,2.5)}$, 
 ${\ll=1}$, 
 MSG: ${\beta=2.8}$, 
 	 ${m=0.76}$, ${\ll=1}$ (${\langle\tilde{r}\rangle=0.519}$), 
 $\phi^4$: ${\lambda=1.0}$, ${m=1}$, ${L=7}$  (${\langle\tilde{r}\rangle=0.5}$)). 
}
\label{fig:4}
\end{figure}

Based on these observations, we analyse how the DSG eigenvector distribution 
depends on ${\ll}$, 
whether it approaches the RMT prediction when moving 
from weak to strong perturbation, and how it changes from one block to another. 
The distributions inside a single block exhibit clearly exponential scaling in the bulk, still with slower decaying tails (Fig.~\ref{fig:3}.g). The slope of this exponential changes from block to block. 
The distribution 
in Fig.~\ref{fig:1}, which corresponds to a large window including many blocks, is actually a superposition of many exponential distributions. 
There is no significant change when ${\ll}$ increases from 0 to ${\sim5}$, the maximum value for which we achieved convergence 
(Fig.~\ref{fig:3}{.h}). In fact, the distribution remains unchanged 
even for larger ${\ll}$, 
where truncation effects are non-negligible. 
At the same time, 
${\langle\tilde{r}\rangle \approx \langle\tilde{r}\rangle_{GOE}}$ 
for any 
${\ll \gtrsim 1}$ and well convergent at least for ${\ll \lesssim 3}$ (Fig.~\ref{fig:3}.f). 

Comparing the eigenvector 
statistics of different models 
in windows with ${\langle\tilde{r}\rangle\approx\langle\tilde{r}\rangle_{GOE}}$, 
we find that they generally vary from model to model but are always different from Gaussian and at best exponential (Fig.~\ref{fig:4}.b). The ${\phi^4}$ model, in particular, deserves special attention. In this case, using the CFT as the unperturbed model to construct the truncation basis is inconvenient, so the massive Klein-Gordon (KG) model ${H_{KG}=H_0 + m^2 U_2}$ is used instead \cite{Hogervorst2014}. 
In contrast to the CFT basis, in KG 
there is no degenerate shell structure. 
Nevertheless, the eigenvector distribution is once again very different from Gaussian and characterised by slowly decaying tails as in DSG. 
Lastly, comparing the single-block eigenvector distributions in SG at different ${\beta}$ (Fig.~\ref{fig:4}.a), we find that they are similar to those of DSG. 
These results clearly show that the discrepancy between the eigenvector distributions and RMT prediction, in particular the presence of slowly decaying tails, is robust under variations of the parameters, energy window, model and truncation basis.

\emph{Discussion.}{---} We have shown that, while the level spacing statistics of 
the above studied non-integrable QFTs 
agree with RMT, their eigenvector component statistics are markedly different from 
RMT predictions. Both of the above features emerge already in the weakly perturbed CFT regime and  
persist unchanged beyond that, which suggests that they may be valid for any perturbation strength. 
Indeed, there is no indication that the scaling of the distributions changes with the perturbation, 
even when the CFT shell structure disappears. 
Moreover, the qualitative characteristics of the eigenvector 
distributions for different models are similar, irrespectively of integrability and even for quite different choices of truncation basis. 
The latter observation particularly rules out an explanation based on exceptional features of the CFT basis. An interesting open question is how the observed discrepancy affects the validity of ETH in (1+1)D QFT. 
Testing ETH using Hamiltonian truncation methods is, however, a more challenging problem, as it is supposed to hold in the thermodynamic limit where the perturbation strength is large and convergence of the spectra worsens. We hope to investigate this question in the future.

The data presented in this work may be accessed at \cite{data}.

\begin{acknowledgments} 
The authors would like to thank Ivan Kukuljan for collaboration and assistance with programming at an early stage of this project. SS would also like to thank Gabor Tak\'acs and Robert Konik for useful discussions. 
MS and SS acknowledge support by the Slovenian Research Agency (ARRS) under grant the QTE (N1-0109). 
TP acknowledges support by the ERC Advanced Grant 694544 -- OMNES and the ARRS research program P1-0402. 
\end{acknowledgments}

\bibliographystyle{apsrev4-2}
\bibliography{quantum_chaos__qft}

\section*{Signatures of Chaos in Non-integrable Models of Quantum Field Theory: \\ Supplemental Material}

In this Supplemental Material we present: 
\begin{itemize}
	\item definitions and detailed analysis of truncation error estimates for the TCSA data,
	\item study of spectral statistics of the DSG in the perturbative regime based on degenerate perturbation theory at the lowest order,
	\item technical information on the implementation of the Hamiltonian Truncation method,
	\item further plots showing the dependence of DSG spectral statistics on the energy in a single shell.
\end{itemize}

\subsection{Truncation Errors}
The main challenge in TCSA is to achieve convergence of the computed quantities for increasing values of the energy cutoff ${E_{\text{cut}}}$. 
Generally, convergence of the numerically computed energy levels is relatively slow (algebraic) but it is possible that
other quantities of interest converge faster than the levels themselves.
{Indeed, TCSA often gives correct results even 
	when the levels have not reached their limiting values, 
	therefore demanding convergence of the 
	spectrum is a rather strict 
	criterion, following which a large amount of useful data would be rejected.}
In the present study, in particular, the statistical properties of spectra and eigenvectors 
are not necessarily sensitive to the precise values of the individual energy levels.
Therefore, it seems reasonable to require convergence of the statistical measures 
instead of the pointwise spectra themselves.

\begin{figure}[htb!]
	\includegraphics[width=1\columnwidth]{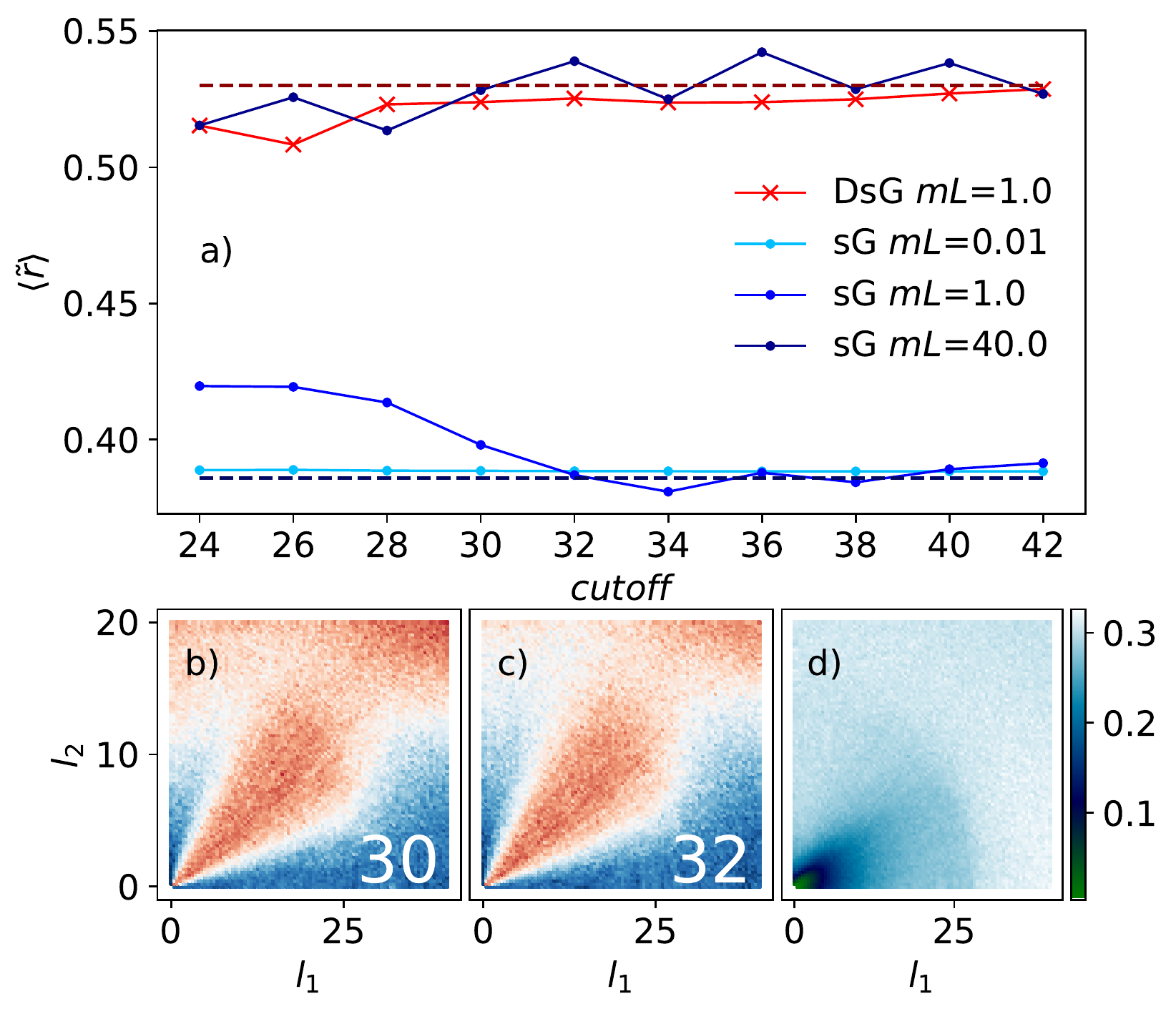}
	\caption{
		Convergence of ${\langle\tilde{r}\rangle}$ for increasing values of the truncation cutoff ${E_\text{cut}/\varepsilon}$. 
		(a) Plots of ${\langle\tilde{r}\rangle}$ as a function of the cutoff 
		in the DSG and SG model for different parameter values (SG: ${\beta=2.5}$, DSG: ${\beta_1=1, \beta_2=2.5}$). The mean ${\langle\tilde{r}\rangle}$ 
		is calculated in the energy window 1300--2000. 
		Integrable sine-Gordon model (SG ${\ll = 40}$) can converge to the RMT value, due to
		non-convergence. 
		The point used throughout the paper (DSG ${\ll = 1}$) is clearly convergent with the cutoff.
		When convergence is reached, the SG spectra are expected to reach Poisson statistics and indeed do so for ${\ll=1, 0.01}$. 
		However for the much larger value ${\ll=40}$ the numerically computed SG spectra show RMT-like behaviour even for the maximum cutoff value 42. This is obviously due to non-convergence of the spectra at this range of cutoff values. 
		Even though the absence of convergence can be seen in the fluctuations of ${\langle\tilde{r}\rangle}$ with the cutoff, 
		such fluctuations cannot be used as a reliable measure of error as they are suppressed when using a larger energy window. 
		(b--c) Density plot of ${\langle\tilde{r}\rangle}$ for the DSG 
		as a function of the two perturbation parameters ${\ll_1,\ll_2}$, 
		at two different cutoffs (${\beta_1=1, \beta_2=2}$, energy window: 2000--3000). 
		When the cutoff increases (here from 30 to 32) the red region in the upper part of the plot moves away, showing that the apparent RMT-like behaviour is due to non-convergence of the spectra. This is an indication that the values in the rest of the plot which remains unchanged are reliable, and we want a measure of truncation error that provides information on whether the computed values are physical or artificial without having to look at the whole parameter space at different cutoffs. 
		(d) The error estimate  ${\langle\Delta\tilde{r}\rangle}$ given by (\ref{eq:error}) calculated from the comparison of data for the cutoffs 30 and 32. \label{fig:appendix_error_cutoff}}
\end{figure}

However, such a criterion may be incorrect and misleading 
when we compare spectral statistics, especially 
${\langle\tilde{r}\rangle}$, with RMT predictions. The problem lies in the
observation that the statistics of a non-convergent spectrum may  
resemble those of random matrices. Statistical measures may 
therefore appear to converge with the cutoff to the RMT predictions 
due to truncation errors instead of physical reasons. In fact, RMT-like statistics
may be observed even for the integrable SG model, as 
${\langle\tilde{r}\rangle}$ may show RMT-like behaviour for relatively low cutoff values,
but switch to Poisson-like behaviour at higher cutoffs. 
This is indeed possible if convergence of the spectrum in the energy
window under study has not been reached at a given cutoff {(Fig.~\ref{fig:appendix_error_cutoff}.a: SG at ${mL=40}$)}.  
{The same can happen also for the non-integrable DSG model 
	at some parameter values (Fig.~\ref{fig:appendix_error_cutoff}.b--c).}
{Conversely, it seems reasonable that spectra whose statistics 
	is Poisson-like have reached convergence and are reliable, since otherwise they would be RMT-like.}

In order to test these statements in an unbiased way, 
we introduce a measure of the truncation error for spectral statistics 
{and check how it varies for increasing cutoff values}. 
The simplest way to estimate the error of the average of a quantity under study at a given cutoff 
is by averaging the absolute change of each individual contribution from one cutoff to the next. 
More explicitly, considering the quantity ${\langle\tilde{r}\rangle}$, we define its error estimate as
\begin{equation}
	\label{eq:error}
	\langle\Delta\tilde{r}\rangle = 
	\frac{1}{N}\sum_{i=0}^{N}
	|\tilde{r}_i^{{(2)}} - \tilde{r}_i^{{(1)}}|,
\end{equation}
where the subscript denotes the index of the ${\tilde{r}}$ value and the superscripts denote the two different energy 
cutoffs ${E_\text{cut}^{(1)}}$ and ${E_\text{cut}^{(2)}}$ with ${E_\text{cut}^{(2)} > E_\text{cut}^{(1)}}$ 
at which the corresponding values are obtained. 
The sum is over the energy window under study, and ${N}$ is the number of
${\tilde{r}}$ values in that window. Based on the assumptions of  TCSA, 
the differences ${\Delta\tilde{r}_i =|\tilde{r}_i^{{(2)}} - \tilde{r}_i^{{(1)}}|}$ should decay 
in the limit of the cutoff going to infinity, and hence the mean value 
${\langle\tilde{r}\rangle}$ is expected to be quite accurate as long as we have reached a sufficiently high cutoff ${E_\text{cut}^{(2)}}$.

\begin{figure}[h!]
	\includegraphics[width=1\columnwidth]{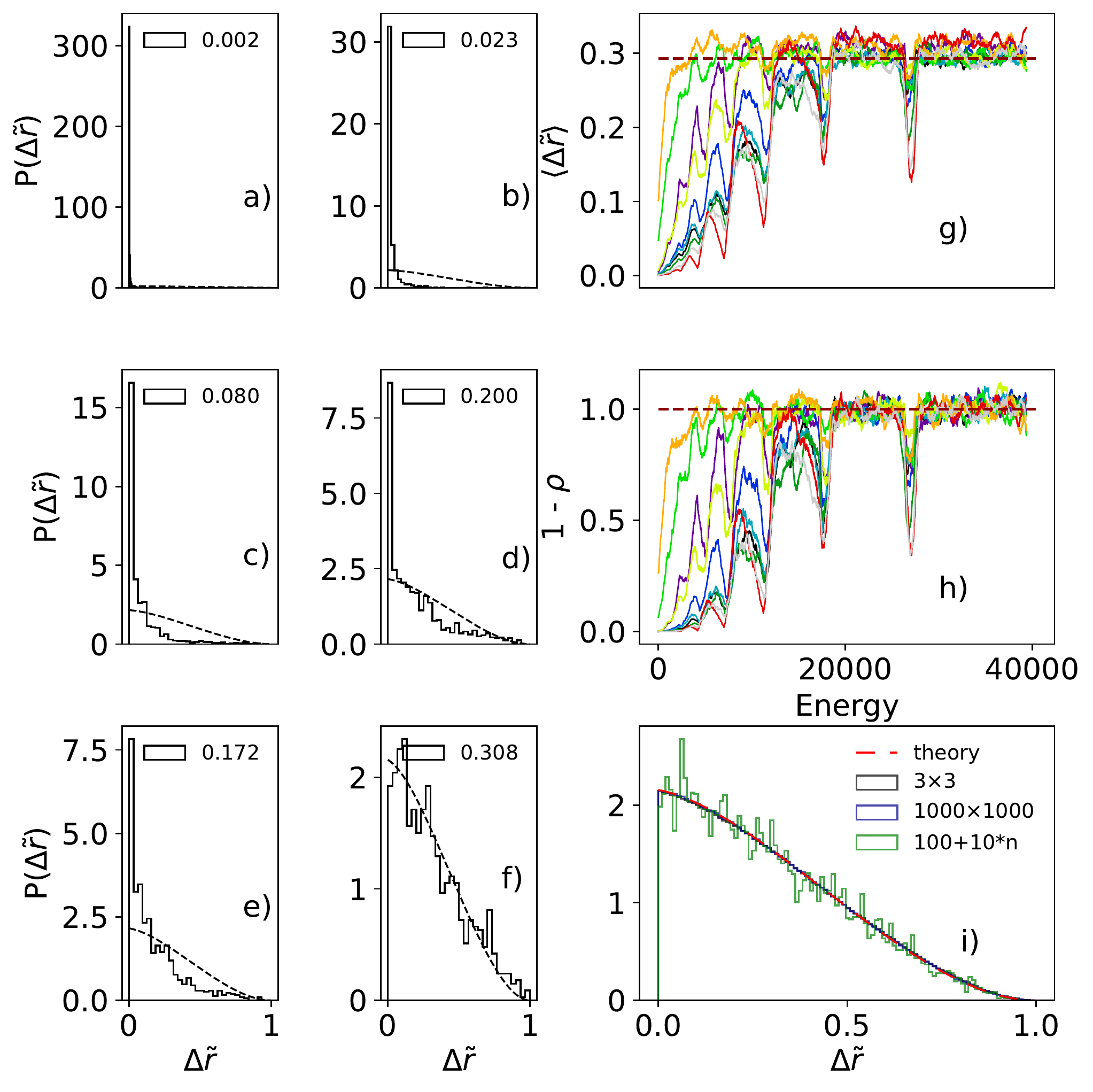}
	\caption{(a--f) The distribution of errors ${\Delta\tilde{r}_i}$ for the SG model in an energy window of 1000 levels at increasing mean energy ($\beta = 0.000011$, ${\ll=1}$). %
		The average error of each distribution is stated in the legend. 
		The black dashed line corresponds to the distribution of ${\Delta\tilde{r}}$ for uncorrelated random matrices. %
		In the case of small average error, the vast majority of the levels change negligibly with the cutoff. 
		At larger average values, the distribution approaches the one of RMT. (g) Moving average of error in spectra
		at randomly chosen different values of the parameters. Window of size 1000. 
		(h) Moving Pearson's
		correlation coefficient for the same spectra as in (g). 
		(i) The probability density distribution of ${\Delta\tilde{r}}$ for ${3\times3}$ random matrices calculated 
		analytically (black dashed), calculated numerically (red) and for ${1000\times1000}$ random matrices
		calculated numerically (blue). %
		The green histogram corresponds to comparison of successive pairs of random matrices of sizes
		${100\times100}$, ${110\times110}$, 
		${120\times120}$, ... up to 1090, such that each one is constructed from the previous one by padding 10 additional rows and columns of random elements, thus resulting in a sequence of not completely uncorrelated matrices. 
		This change is analogous to a cutoff increase in TCSA, showing that even a relatively small change in a random matrix results in a large error estimate, thus demonstrating that ${\langle\Delta\tilde{r}\rangle}$ is a very sensitive measure of convergence. 
	}
	\label{fig:correlation_versus_error}
\end{figure}

But there is more to this error estimate, which makes it very reliable. First, 
${\langle\Delta\tilde{r}\rangle}$ is bounded, taking its maximum value
when the two spectra under comparison correspond to spectra of independent random matrices.
Second, it behaves analogously to the measure of correlation between the two compared spectra (Fig.~\ref{fig:correlation_versus_error}.g--h). 
More specifically, when ${\langle\Delta\tilde{r}\rangle}$ increases then the quantity ${1-\rho}$, 
where ${\rho}$ is the Pearson coefficient of correlation between the two spectra, also increases. 
If the spectra change with the cutoff so much that they are not correlated at all,  
we can say that they are maximally non-convergent. We can then estimate the maximum 
error by calculating the average error between two uncorrelated random matrices.
This can be performed analytically for ${3\times3}$ matrices by deriving the probability density
of the error 
\begin{equation}
	P(\Delta\tilde{r}) = \int_{0}^{1}\int_{0}^{1}P(\tilde{r}_1)
	P(\tilde{r}_2)\delta(|\tilde{r}_1-\tilde{r}_2|-\Delta\tilde{r})
	\mathrm{d}\tilde{r}_1 \mathrm{d}\tilde{r}_2
\end{equation}
and finding its mean. The average error can then be calculated to be {$\langle\Delta
	\tilde{r}\rangle_{\mathrm{RMT}} \approx 0.29$}. %
We found numerically that this distribution is very general
and holds for matrices of sizes at least up to ${1000\times1000}$ and
even for spectra of matrices that differ only by a few additional rows 
(Fig. \ref{fig:correlation_versus_error}.i).

In the models under study we indeed observe that the error estimate is 
maximal for the least convergent spectra (Fig.~\ref{fig:correlation_versus_error}.g--h). 
Using this error we can now distinguish between spectra that exhibit RMT-like spectral statistics due to physical reasons or due to truncation artefacts. If the average error is close to the maximum value corresponding to the difference between two uncorrelated random matrices, then the spectra must be random due to non-convergence.
If instead we observe very small average errors of the order of, say, ${10\%}$ of the maximum value, then 
we can trust that the randomness is due to genuine physical reasons.
It should be noted that this measure is still quite conservative, because, 
as the ${r}$ values are computed from differences of consecutive levels,
they are very sensitive to truncation errors and converge much slower than the absolute spectra themselves.
Moreover, passing from one cutoff to the next the ordering of a non-negligible number of levels 
changes, which means that the corresponding differences ${\Delta r_i}$ are larger than the actual values 
resulting in an overestimation of the error.

\begin{figure}[h!]
	\includegraphics[width=1\columnwidth]{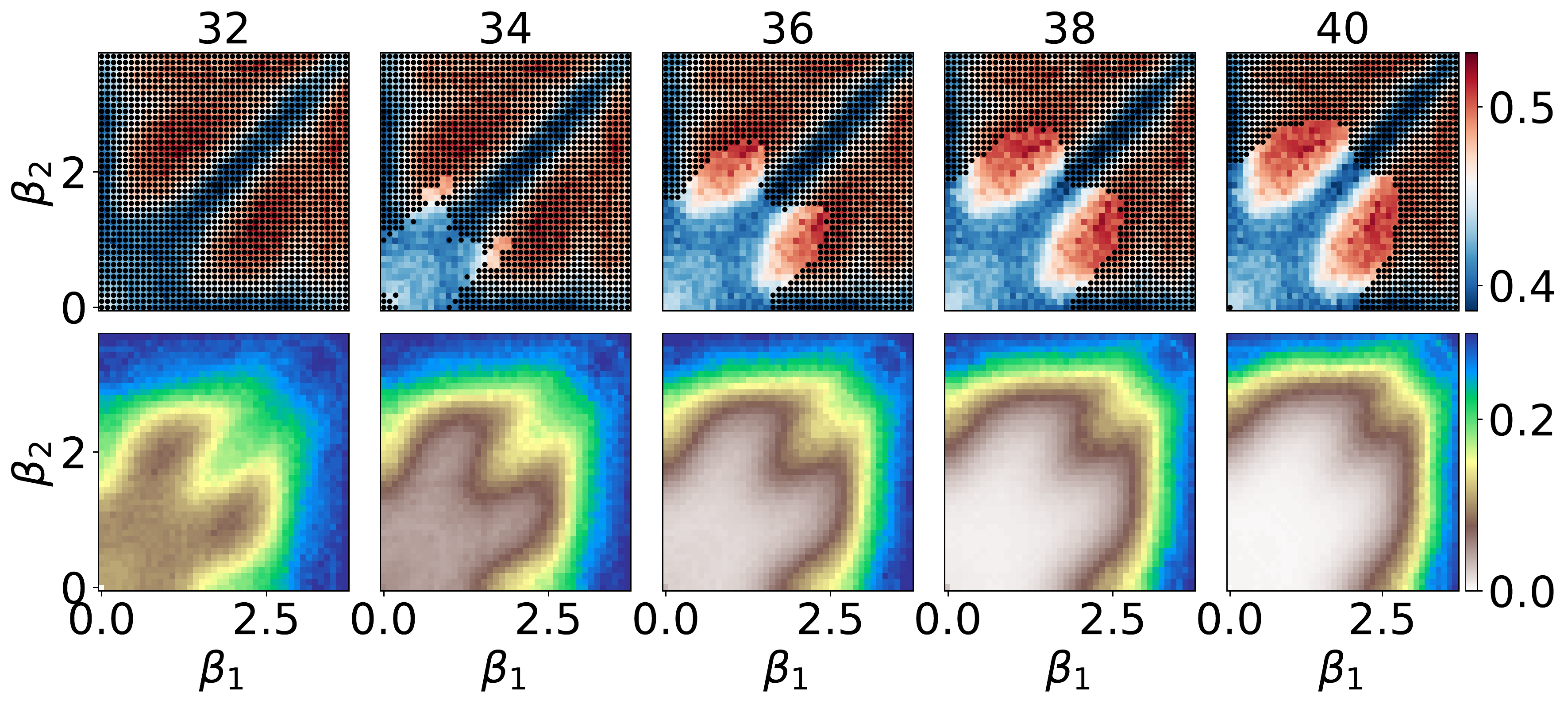}
	\caption{Density plots of ${\langle\tilde{r}\rangle}$ for the DSG model as a function of the parameters ${\beta_1,\beta_2}$; same as in Fig.~\ref{fig:2} of the main text, but showing also the estimated truncation error. \emph{Top}: Plots for increasing cutoffs, with the black dots indicating the region where the error ${\langle\Delta\tilde{r}\rangle}$ is larger than 0.05. \emph{Bottom}: Density plots of the error estimate ${\langle\Delta\tilde{r}\rangle}$. 
		We observe that as the cutoff increases the error decreases at every point in parameter space and the area where the error is smaller than the 0.05 threshold expands from smaller to larger ${\beta_1,\beta_2}$.}
	\label{fig:density_plots+errors}
\end{figure}

{Equipped with a reliable truncation error estimate, 
	we can now test quantitatively the quality of TCSA convergence 
	by analysing how the error changes for increasing values of the cutoff (Fig.~\ref{fig:density_plots+errors}).  %
	We generally find that, when the cutoff goes beyond a sufficiently large value, 
	${\Delta\tilde{r}}$ starts decreasing from the maximum value that it has at lower cutoffs 
	and keeps decreasing while at the same time fluctuating as the cutoff increases further.
	Based on this convergence tests we find that the best convergence for ${\langle\tilde{r}\rangle}$ 
	is achieved in the area around the lower left corner of the phase diagram shown in Fig.~\ref{fig:2}. 
	Moreover, we find that ${\langle\tilde{r}\rangle}$ converges quite well 
	for ${\ll\lesssim 3}$ in the DSG case and ${\ll\lesssim 1}$ in the SG case. 
	Lastly, we verify that convergence is achieved first at lower energy windows (Fig.~\ref{fig:correlation_versus_error}.g--h) 
	with the lowest 3000 to 4000 levels corresponding to well convergent ${r}$ values at ${\ll\lesssim 1}$ 
	in the parameter range of Fig.~\ref{fig:2}. 
	The spectra and eigenvectors presented in the plots of this paper are 
	sufficiently well convergent with average errors of at most ${\langle\Delta\tilde{r}\rangle = 0.05}$ and values of ${\langle\tilde{r}\rangle}$ that do not change significantly with the cutoff, unless otherwise stated (Fig.~\ref{fig:3}
	,  ${\ll\gtrsim 5}$). %
}

The proposed error measure improves considerably 
our ability to confidently distinguish physical from numerically artificial results. 
We now realise that the intuitive guess that 
non-convergent spectra are RMT-like is not always correct. 
Even though we do observe that this behaviour is the typical case, 
we can also clearly find exceptions to this rule: 
Poisson-like spectral statistics can be observed even 
in the region with maximal non-convergence 
(Fig.~\ref{fig:appendix_error_cutoff}.c). 

{While we have so far focused on ${r}$ values, we can estimate in a similar way the convergence of eigenvector statistics  by comparing how much each eigenvector changes from cutoff to cutoff. 
	We empirically found that a reliable test for the comparison of two vectors 
	can be based on the correlation between their cumulative squared coefficients, 
	which can be used as the fingerprint of a vector. 
	More explicitly, for each numerically computed vector ${|\Phi_{j}\rangle}$ we construct the list 
	${\sum_{k=1}^i |c_{kj}|^2}$ for ${i=1,\dots, N}$ where ${N}$ is the length of the shortest between the two compared vectors, 
	and then compare the two lists by computing the Pearson correlation coefficient. 
	It should be noted in passing that this comparison function is also useful for tracking the eigenvectors as we incrementally change a parameter (physical parameter of the Hamiltonian or the cutoff) and identifying changes in the energy level ordering between two consecutive steps. To this end, we pairwise compare all eigenvectors of one matrix with those of the other and match them based on which pairs show the highest correlation. Typically, such ordering changes are limited to first or second neighbours. In this way we can correct the order of the eigenvectors and therefore obtain a more accurate estimate of the convergence error. }

{Using this error estimate we found that eigenvector statistics are convergent for a wider range of parameter values than ${\tilde r}$ values. 
	This can be easily understood from the fact that the ${\tilde r}$ values are more sensitive to small changes in the energy level values.
	We found that, at the maximum cutoff ${E_\text{cut}/\varepsilon=42}$ used here and for the values of ${\beta_i}$ as in Fig.~\ref{fig:3}
	, eigenvector statistics are well convergent for up to ${\sim3000}$ levels at ${\ll=5}$.}

\subsection{Quantum chaos and perturbed CFT}
\label{subsec: Quantum chaos and CFT}

As shown in the main text, the emergence of chaotic level spacing statistics in the DSG occurs already at infinitesimal values of the perturbation strength, as long as the two cosine perturbations are mixed with approximately equal coefficients (Fig.~\ref{fig:2b}.a). 
This observation motivates us to study the problem using first order perturbation theory. Like any CFT, the spectrum of the free massless boson field theory, which is the unperturbed model, exhibits extensive degeneracies, in which case perturbation theory tells us that at first order the perturbation affects each of the degenerate energy shells independently from the others. This means that at first perturbative order we can analyse the spectral statistics working on finitely dimensional Hilbert spaces (those of the degenerate energy shells), that is, without having to deal with the problems arising from an infinite Hilbert space dimension and the need to truncate it in an efficient way. This is especially convenient as it allows us to derive exact spectra and statistics,  reach higher energy shells and study the changes in spectral statistics for increasing shell number. In this appendix, we present more detailed results for the statistics of level spacings and eigenvectors of the DSG model at first order in perturbation theory.

The Hamiltonian of the free massless boson field theory with Dirichlet boundary conditions is 
(up to an irrelevant additive constant)
\begin{equation}
	H_0 = \sum_{n=1}^\infty E_n a_n^\dagger a_n 
\end{equation}
where ${a_n, a_n^\dagger}$ are the ladder operators corresponding to the sinusoidal harmonic eigenfunctions ${\sqrt{2/L}\sin(k_n x)}$ with wavenumbers ${k_n= n\pi/L, \; n=1,2,\dots}$, satisfying canonical commutation relations ${[a_n,a_{n'}^\dagger] =\delta_{n,n'}}$, and the single particle dispersion relation is ${E_n=k_n}$ (in units where the speed of light is ${c=1}$ and ${\hbar=1}$). 
The eigenstates of ${H_0}$ are therefore
\begin{equation}
	|\{\nu_n\}\rangle = \prod_{n=1}^\infty \frac{(a_n^\dagger)^{\nu_n}}{\sqrt{\nu_n !}}|0\rangle
\end{equation}
with eigenvalues
\begin{equation}
	E(\{\nu_n\}) = \frac{\pi}{L} \sum_{n=1}^\infty n{\nu_n} 
\end{equation}
Evidently, owing to the linearity of the dispersion relation, all energy eigenvalues of ${H_0}$ are integer multiples of ${\pi/L}$ and can be classified into shells of degenerate levels with energy ${E=N\pi/L}$ for all positive integers ${N}$ and degeneracy equal to the number ${P(N)}$ of integer partitions of ${N}$. For large ${N}$ the number ${P(N)}$ increases rapidly (${\log(P(N))\propto \sqrt{N} }$). For this reason the dimensions of CFT degenerate shells increase with the energy faster than in few-body quantum mechanical systems (e.g. hydrogen atom). We will denote the subspace corresponding to the shell with energy ${E=N\pi/L}$ as ${\mathcal{H}_N}$. Taking into account the additional restriction to a single symmetry sector (see Sec.~\ref{sec:Details}), the resulting shell sizes are listed in Tab.~\ref{tab:shell_sizes}.

\begin{table}
	\begin{tabular}{ |r|r|r|  }
		\hline
		shell no. & ${\qquad N}$ & number of states \\ 
		\hline
		10 & 22 & 505 \\
		11 & 24 & 793 \\
		12 & 26 & 1224 \\
		13 & 28 & 1867 \\
		14 & 30 & 2811 \\
		15 & 32 & 4186 \\
		16 & 34 & 6168 \\
		17 & 36 & 9005 \\
		\hline
	\end{tabular}
	\caption{Number of states of the degenerate energy shells.}	
	\label{tab:shell_sizes}
\end{table}

Let us consider a perturbation of ${H_0}$ by an operator ${V}$ that raises the degeneracy of the energy levels at first order in the perturbation parameter ${\lambda}$. This is true for the DSG interaction. From degenerate perturbation theory we know that the perturbed energy eigenstates of ${H=H_0+\lambda V}$ corresponding to the shell ${\mathcal{H}_N}$ are 
\begin{equation}
	|\Phi_{N,j}\rangle = \sum_{j=1}^{P(N)} c_{i,j}|\Phi_{N,i}^0\rangle + \mathcal{O}(\lambda)
\end{equation}
with energy eigenvalues
\begin{equation}
	E_{N,j} = E_{N} + \lambda E^{(1)}_{N,j} + \mathcal{O}(\lambda^2)
\end{equation}
where ${|\Phi_{N,i}^{(0)}\rangle \equiv \sum_{j=1}^{P(N)} c_{i,j}|\Phi_{N,i}^0\rangle}$ and ${E^{(1)}_{N,j}}$ are the eigenvectors and corresponding eigenvalues of the restriction ${V_N}$ of ${V}$ in the subspace ${\mathcal{H}_N}$. 
In the limit ${\lambda\to 0}$ the perturbed eigenstates are linear combinations of states of the shell they originate from and only those. We will therefore focus on the eigenvectors ${|\Phi_{N,i}^{(0)}\rangle}$ and eigenvalues ${E^{(1)}_{N,j}}$ of ${V_N}$ whose statistics are identical to those of ${H}$ in the subspace ${\mathcal{H}_N}$ as ${\lambda\to 0}$. In the case of the DSG model, ${\lambda V=\lambda_1 V_1 + \lambda_2 V_2}$ therefore in the above limit the spectrum is a function of the mixing parameter ${x=\lambda_1/(\lambda_1+\lambda_2)}$. For ${x=0}$ or ${1}$ the model reduces to the SG.

\begin{figure}[htbp]
	\includegraphics[width=1\columnwidth]{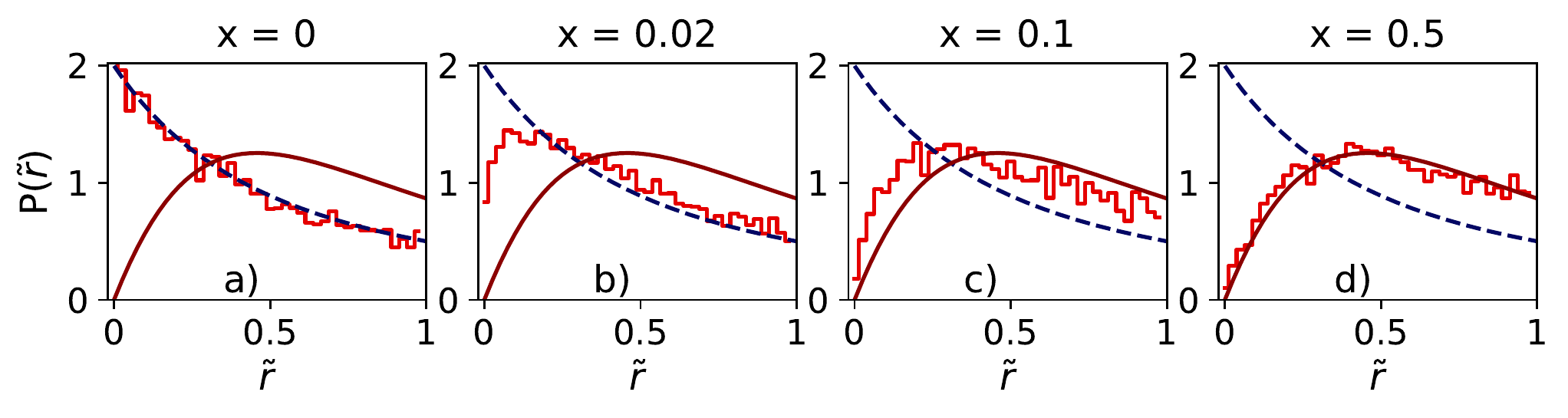} 
	\caption{Histograms of ${\tilde{r}}$ values for the DSG in-shell spectrum for four different values of the mixing parameter ${x}$ showing the change from Poisson at ${x=0}$ to GOE statistics at ${x=0.5}$. The histograms are computed from spectra of the shell ${N=36}$ (9005 levels).}
	\label{fig:rthist}
\end{figure}
\begin{figure}[htbp]
	\includegraphics[width=1\columnwidth]{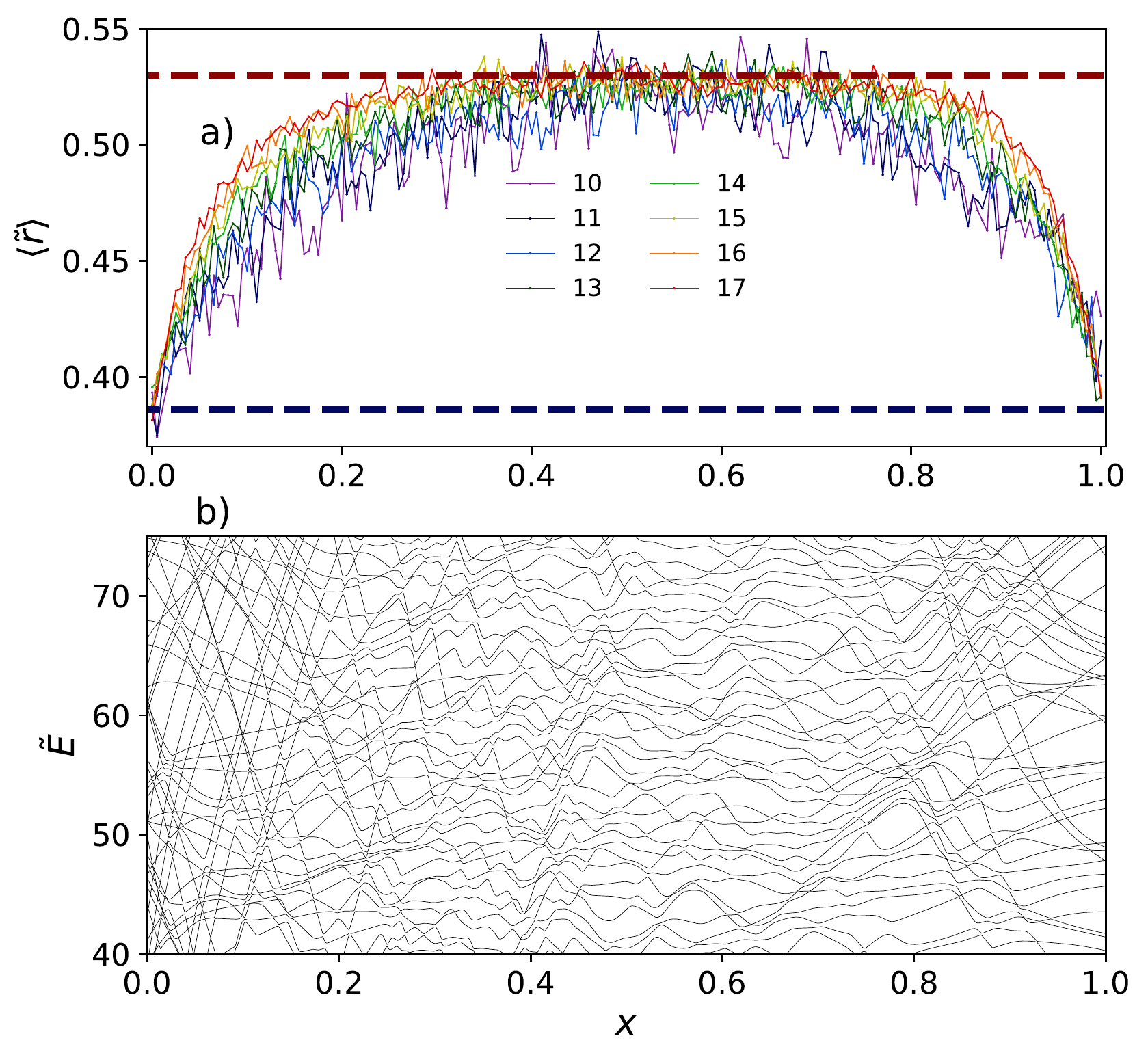}
	\caption{(a) Variation of ${\langle\tilde{r}\rangle}$ of the DSG in-shell spectra as a function of the parameter ${x}$ of mixing of the two cosine perturbations. The curves correspond to the shells  with ${N=22,24,26,\dots,36}$ (10\textsuperscript{th} to 17\textsuperscript{th} shell). (b) A sample of DSG energy levels for varying mixing parameter ${x}$ from 0~to~1. We observe the level repulsion for most values of ${x}$ except close to the edges where the DSG reduces to the SG model. The sample corresponds to the 10\textsuperscript{th} energy shell and the unfolding procedure has been applied to uniform the level density.}
	\label{fig:rtmean}
\end{figure}

\begin{figure*}[htbp]
	\includegraphics[width=0.9\textwidth]{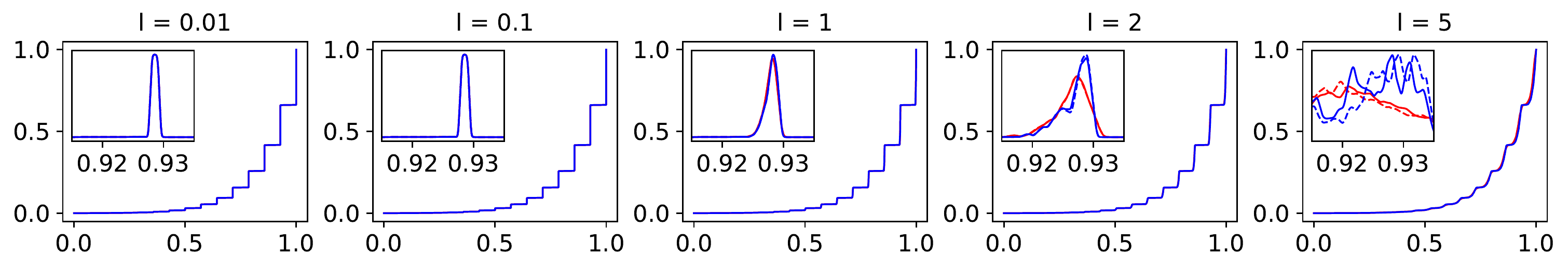}
	\caption{The cumulative spectral density of DSG (red lines) and SG (blue lines) at different values of ${\ll=\ll_1=\ll_2}$. The spectra correspond to the lowest 15 energy shells (5173 levels), rescaled from 0 to 1 for easier comparison. The insets show the spectral density in a single shell (${N=26}$, levels: 2082--3305). The dashed lines correspond to the spectra of lowest-order perturbation theory. The staircase form, which originates from the characteristic shell structure of CFT spectra, is present for all ${\ll  \lesssim 1}$ for both the DSG and SG spectra, whose densities are hardly distinguishable. For larger ${\ll}$ the two spectral densities gradually become visibly different from each other and from the perturbative spectra.
	}
	\label{fig:spectra}
\end{figure*}

In the following we will analyse the dependence of spectral statistics 
on the mixing parameter ${x}$ and shell number ${N}$, keeping ${\beta_1}$ and ${\beta_2}$ fixed at the values used in the main text, i.e. 1 and 2.5 respectively. %
Fig.~\ref{fig:rthist} shows histograms of ${\tilde r}$ based on in-shell spectra {for the 17\textsuperscript{th} shell (${N=36}$)} at various values of the mixing parameter, demonstrating the transition from Poisson statistics at ${x=0}$ to GOE at ${x\approx1/2}$ where ${{\langle\tilde{r}\rangle}}$ is maximal. Owing to the larger spectrum size, the statistical fluctuations of the histograms are reduced compared to Fig.~\ref{fig:1} 
of the main text, allowing a more accurate comparison with the theoretical curves. 
In Fig.~\ref{fig:rtmean}.a showing ${{\langle\tilde{r}\rangle}}$ as a function of ${x}$ in different shells, it is evident that as we move from lower to higher shells, ${{\langle\tilde{r}\rangle}}$ approaches the GOE value closer and in a broader window of ${x}$ values expanding from the middle ${x\approx1/2}$. This suggests that  at higher energy levels the RMT behaviour of level spacings not only persists but also becomes more prominent. 
In-shell eigenvector statistics exhibit identical scaling as that shown in Fig.~\ref{fig:3} 
in the main text, which is computed from spectra at intermediate values of the perturbation strength.

As pointed out in the main text, the above general features of spectra and eigenvectors observed in the limit ${\lambda\to 0}$ persist for all values of the perturbation parameters for which convergence was achieved in our study. Fig.~\ref{fig:spectra} shows that the spectra exhibit a distinct shell structure characteristic of first-order perturbation theory for all ${\ll\lesssim1}$, while the shells start sensing each other at ${\ll\approx 1}$ and visibly mix with each other at larger ${\ll}$. In particular, at the maximum value ${\ll=5}$ for which we have achieved convergence of the spectrum and eigenvectors (lowest ${{\sim}\,3000}$ levels), the effects of the interaction are strong enough to result in a spectral density that is clearly different from that of first-order perturbation theory. Yet the level spacing and eigenvector statistics for ${\ll=5}$ exhibit the same characteristics as for ${\ll\to 0}$.

\subsection{Details of the Hamiltonian Truncation Method}
\label{sec:Details}

{As mentioned in the main text, Hamiltonian truncation methods have been extensively used
	for the derivation of the spectra of almost all of the models studied in the present work. 
	Details on the application of TCSA to SG and DSG can be found in \cite{TCSA-sG1,TCSA-sG2,TCSA-sG3,KST} 
	and \cite{Bajnok_2001,Tak_cs_2006} respectively, 
	while the application to MSG is a relatively straightforward extension of the method. 
	Details on Hamiltonian truncation in the ${\phi^4}$ model can be found 
	in \cite{Hogervorst2014,Rychkov2014,Rychkov2015,Elias-Miro2017}. 
	Here we provide further information relevant for the analysis of spectral statistics in these models.  
	More specifically, we discuss the discrete symmetries of the models and restriction to a single symmetry sector, provide information on the size of the truncated bases we used and hints on the efficient construction of the Hamiltonian matrices. 
	Lastly, we give explicit formulas for the parameters of the models as used here.
}

\subsubsection*{Massless free boson basis}

We express the SG, DSG and MSG Hamiltonians in the truncated
free boson CFT eigenstate basis ordering the states by their energy. 
We assume {Dirichlet boundary conditions} in a box of length ${L}$. 
Both the unperturbed and perturbed Hamiltonians 
are invariant under two discrete ${\mathbb{Z}_2}$ transformations: 
field reflection ${\phi\to-\phi}$ and space reflection ${x\to L-x}$. 
For the study of spectral statistics it is necessary to eliminate any symmetries of the model
by restricting the spectrum to a single symmetry sector 
(energy levels belonging to different symmetry sectors are independent and can clearly cross with each other).
Out of the four symmetry sectors we choose the one containing the ground state. 
The truncated basis sizes of the selected symmetry sector are listed in Tab.~\ref{tab:basis_sizes}. 
The Hamiltonian matrix is dense in this basis and to derive the spectra we use exact diagonalisation.
The construction time of the  Hamiltonian matrix can be  reduced considerably using simple parallelisation.

\begin{table}
	\begin{tabular}{ |r||r|r|  } 
		\hline
		cutoff & number of states & CPU time [days]\\
		\hline
		32 & 12170 & 0.7\\
		34 & 18338 & 2\\
		36 & 27343 & 4\\
		38 & 40369 & 8\\
		40 & 59061 & 17\\
		42 & 85674 & 36\\
		\hline
	\end{tabular}
	\caption{Truncated basis sizes and corresponding computing time for the construction of the Hamiltonian matrices for the cutoff values ${E_{\text{cut}}/\varepsilon}$ used in the present study.}	
	\label{tab:basis_sizes}
\end{table}

\subsubsection*{Klein Gordon basis}

To achieve sufficiently good convergence 
the ${\phi^4}$ Hamiltonian should be expressed in the KG instead of the massless free boson basis \cite{Hogervorst2014,Bajnok2016}. 
The perturbation operator ${U_4}$ (more generally any ${U_n}$ operator) is sparse in this basis, 
and so the construction of the corresponding matrix can be programmed efficiently with the use of dictionaries 
associating each basis state with its order in the basis. 
This way we can benefit from the logarithmic lookup time and considerably speed up the calculation. 
For easier comparison with earlier literature \cite{Hogervorst2014,Bajnok2016}
we focused on the case of periodic boundary conditions. 
This means that  we now have translational invariance symmetry, 
apart from the ${\mathbb{Z}_2}$ symmetry under field reflection ${{\phi\to-\phi}}$. 
We again restrict ourselves to a single symmetry sector, the one containing the ground state. 
For the basis truncation we set the total 
{momentum cutoff} %
at 42 and the system size is chosen to be ${mL=7}$, 
where ${m}$ is the KG mass parameter. 
This choice corresponds to a basis size of ${1{,}504{,}767}$. 
To compute the spectra and eigenstates of the ${\phi^4}$ Hamiltonian, 
we employed sparse matrix diagonalisation techniques, which allowed us
to obtain the lowest ${{\sim}\,500}$ eigenvalues and eigenstates. %

\subsubsection*{Construction of the SG, DSG and MSG Hamiltonian matrices}

The free boson CFT Hamiltonian for a system of length ${L}$ 
with Dirichlet boundary conditions, expressed in dimensionless form in units of a mass scale ${M}$, is 
\begin{equation}
	H_0/M = \frac{\pi}{\ll}\sum_{n=1}^\infty n a_n^\dagger a_n
\end{equation}
where ${\ll=ML}$. 
The sine-Gordon Hamiltonian is defined as
\begin{equation}
	H_{\mathrm{SG}}=H_0+ \lambda V_\beta
\end{equation}
with
\begin{align}
	V_\beta & = - \int_0^L \cos\beta\phi(x) \mathrm{d}x \nonumber \\
	& = 
	- \frac{\pi}{L} \frac12 \int_0^L \left( \mathcal{V}_{+\beta}(x, t) + \mathcal{V}_{-\beta}(x, t) \right) \mathrm{d}x
\end{align}
where 
\begin{equation}
	\mathcal{V}_{\beta}(x, t)= %
	\mathrm{e}^{\mathrm{i}\beta\phi(x)}
\end{equation}
is the vertex operator. 
As in the standard TCSA notation, the perturbation strength ${\lambda}$ has been re-parametrised in favour of the 
first sine-Gordon breather mass in units of the inverse system size ${\ll=m_\beta L}$, 
which is equal to the energy gap between the ground and first excited state in the thermodynamic limit
\begin{equation}
	\lambda(\beta,\ll) = \frac{\kappa(p(\beta))}{2}\Big(\frac{\pi}{\ll}\Big)^{\frac{p(\beta)-1}{p(\beta)+1}}
\end{equation}
where 
\begin{align}
	\kappa(p)&=\frac{2}{\pi}\frac{\Gamma\left(\frac{p}{p+1}\right)}{\Gamma\left(\frac{1}{p+1}\right)}\left[\frac{\sqrt{\pi}\Gamma\left(\frac{p+1}{2}\right)}{2\Gamma\left(\frac{p}{2}\right)}\right]^{2/(p+1)},
	\\
	p(\beta) &= \frac{\beta^{2}}{8\pi - \beta^{2}},
\end{align}
and 
\begin{equation}
	\label{eq:breather mass}
	{m_\beta} = 2M\sin(\pi p/2)
\end{equation}
is the first breather mass, where ${M}$ is the soliton mass.

To construct the double sine-Gordon Hamiltonian with frequencies ${\beta_1}$ and ${\beta_2}$, 
we add the corresponding sine-Gordon Hamiltonians. More explicitly, when we vary ${\beta_{1,2}}$ 
we adjust the coefficients so as to keep the values of ${\ll_{1}, \ll_{2}}$ fixed and equal ${\ll_{1}=\ll_{2}=\ll}$, and compute the spectra of the dimensionless Hamiltonian matrix
\begin{align*}
	H_{\mathrm{DSG}} = 2H_0(\ll) + \lambda(\beta_1,\ll) V_{\beta_1} + \lambda(\beta_2,\ll) V_{\beta_2}
\end{align*}
When we vary ${\ll_{1}}$ and ${\ll_{2}}$ at fixed ${\beta_{1}}$ and 
${\beta_{2}}$ (Fig. \ref{fig:appendix_error_cutoff}), we use the Hamiltonian matrix
\begin{align*}
	H_{\mathrm{DSG}} & =  \ll_1 H_0(\ll_1) + \ll_2 H_0(\ll_2) \\
	& \qquad + \ll_1 \lambda(\beta_1,\ll_1) V_{\beta_1} + \ll_2 \lambda(\beta_2,\ll_2) V_{\beta_2} 
\end{align*}
When studying the massive Schwinger-Thirring model, we write the Hamiltonian as
\begin{align*}
	H_{\mathrm{MSG}} =  \ll H_0(\ll)  + \ll \lambda(\beta,\ll) V_{\beta} 
	+\frac{m^{2}}{2}\int_{0}^{L}:\phi^{2}:dx.
\end{align*}

\subsection{DSG spectral statistics in subsets of states within one shell}

\begin{figure}[htbp]
	\includegraphics[width=\columnwidth]{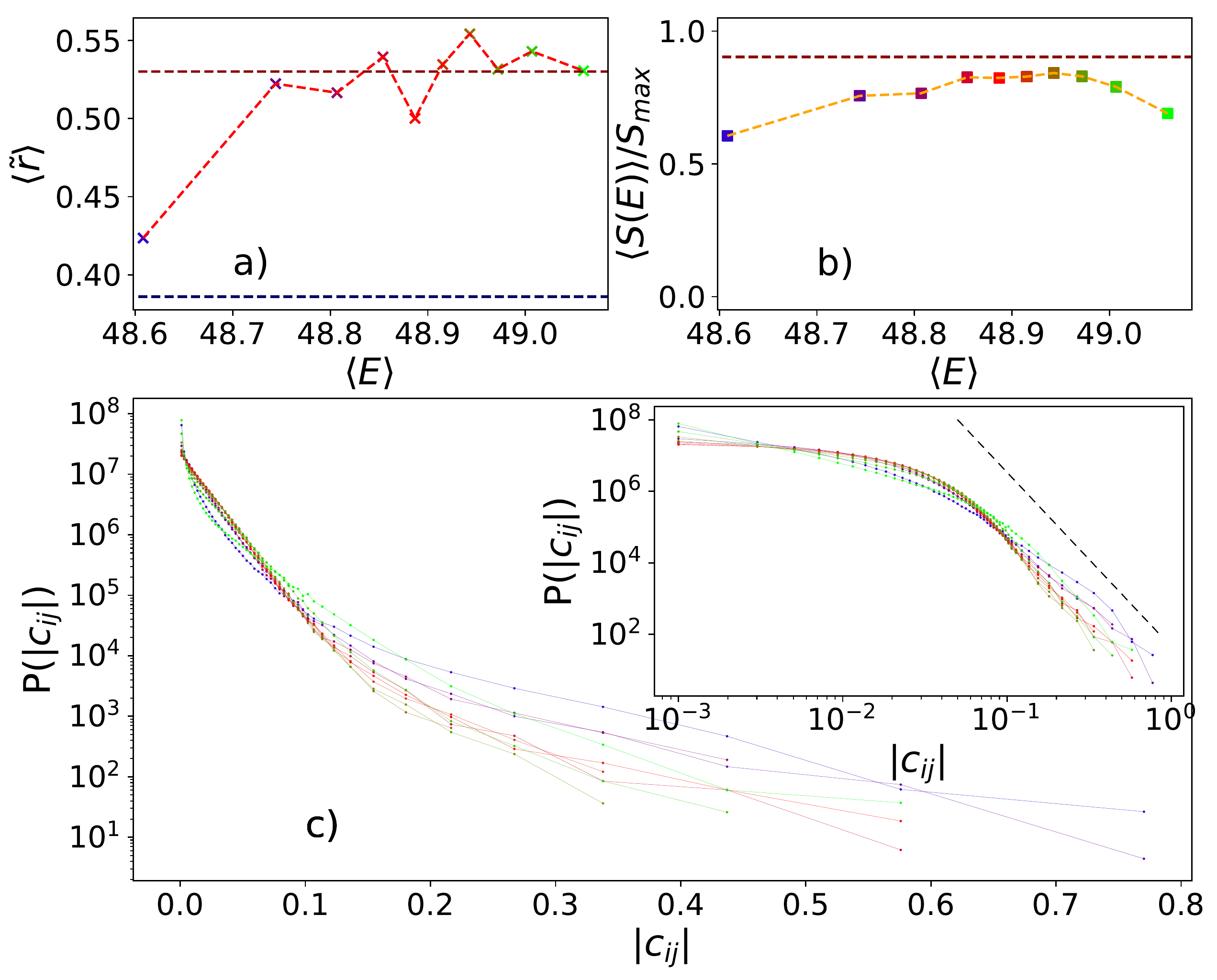}
	\caption{
		Spectral statistics of the DSG model in subsets of a single energy shell. (a--b) Mean ${\langle\tilde{r}\rangle}$ and mean entropy ${\langle S(E)\rangle/S_\text{max}}$ in each subset as a function of the mean energy ${\langle E\rangle}$ of the subset. The dashed line in (b) corresponds to the GOE value ${\langle S\rangle_\text{GOE}=\log(0.48 D)}$ for large $D$. 
		(c) Distribution of ${|c_{ij}|}$ in each subset in log-plot and log-log-plot in the inset (colors as in (a--b) above). 
		Despite the variations of statistics when moving from the edges to the middle of the shell, the tails of the eigenvector component distributions are always different from Gaussian. 
		Each subset consists of 186 consecutive levels of the 13\textsuperscript{th} shell (${N=28}$, corresponding roughly to the same energy window as in Fig.~\ref{fig:1} and \ref{fig:2}
		) and the parameter values are ${\ll=1, (\beta_1,\beta_2) = (1.0,2.5)}$. 
	}
	\label{fig:extra_fig}
\end{figure}

To investigate the inner structure of the energy shells of the DSG model presented in the main text (Fig.~\ref{fig:3}
), 
we focus on a single shell, divide it into non-overlapping subsets of increasing energy levels and analyse the statistics of eigenvector components in each of the subsets. This way we can check whether the observed deviation of the eigenvector component statistics from RMT is due to the existence of a small number of non-typical eigenstates concentrated at the edges of the shell or if it is a general characteristic of the majority of the eigenstates. As a measure of randomness of the eigenvectors ${|\Phi_{j}\rangle}$, we also compute the entropy ${S_j = -\sum_{i} |c_{ij}|^2 \log|c_{ij}|^2}$. The maximum value of the entropy ${S_\text{max}=\log({D})}$ corresponds to a vector with all components equal (where $D$ is the length of the vector). The expectation value of the entropy of a random vector in GOE is ${\langle S\rangle_\text{GOE}=\log(0.48 D)}$ for large $D$ \cite{Kota}. 

The results are shown in Fig.~\ref{fig:extra_fig}. We observe that the eigenstates close to the edges of the shell exhibit less random characteristics. The mean eigenvector entropy ${S(E)}$ is lower than in the middle of the shell, and also ${\langle\tilde{r}\rangle}$ has a lower value at the low energy edge. Moreover, there are variations in the distribution of eigenvector components. In contrast to the middle of the shell, the edges are characterised by non-exponential scaling in the bulk of the distribution and a higher peak at zero. However, the tails of the distributions are clearly non-Gaussian in all subsets independently of their position in the shell. We have verified that the same holds for different parameter values and different shells. We thus conclude that the deviations from RMT predictions are present in the full energy range of each shell, and they are not caused by a small number of non-typical eigenstates.


\clearpage

\end{document}